# Synthesis, Optical, and Magnetic Properties of $Ba_2Ni_3F_{10}$ Nanowires

Shuang Zhou[†], Ji Wang[‡, §], Yakui Weng[†], Zhangting Wu[†], Zhenhua Ni[†], Qingyu Xu[*, †, ‡], Jun Du[*, ‡, §] and Shuai Dong[*†]

[†] Department of Physics, Southeast University, Nanjing 211189, China

[‡] National Laboratory of Solid State Microstructures, Nanjing University, Nanjing 210093, China

[§] Department of Physics and Collaborative Innovation Center of Advanced Microstructures, Nanjing University, Nanjing 210093, China

**ABSTRACT:** A low temperature hydrothermal route has been developed, and pure phase $Ba_2Ni_3F_{10}$ nanowires have been successfully prepared under the optimized conditions. Under the 325 nm excitation, the $Ba_2Ni_3F_{10}$ nanowires exhibit three emission bands with peak positions locating at 360 nm, 530 nm, and 700 nm, respectively. Combined with the first-principles calculations, the photoluminescence property can be explained by the electron transitions between the $t_{2g}$ and $e_g$ orbitals. Clear hysteresis loops observed below the temperature of 60 K demonstrates the weak ferromagnetism in $Ba_2Ni_3F_{10}$ nanowires, which has been attributed to the surface strain of nanowires. Exchange bias with blocking temperature of 55 K has been observed, which originates from the magnetization pinning under the cooling field due to antiferromagnetic core/weak ferromagnetic shell structure of $Ba_2Ni_3F_{10}$ nanowires.

**Keywords:** fluorides, nanowire, photoluminescence, exchange bias, multiferroic.

# 1. INTRODUCTION

Fluorides have attracted considerable research interests for their optical properties, especially the photoluminescence (PL) which has been successfully applied in the phosphor-converted white light-emitting diodes.[1,2] Recently, partial fluorides containing the octahedra have also been considered as potential candidates for the single-phase multiferroic materials with the simultaneous coexistence of magnetism and ferroelectricity.[3] Due to the higher electronegativity of F compared with O, the $M$-F ($M$ denotes metallic ion) bond is more ionic than the $M$-O one. Thus weaker hybridization can be expected between $M$ and F, and no significant charge transfer occurs in the $M$-F bond which is common in oxide ferroelectrics.[4] This indicates that the ferroelectricity in fluorides is of different origin than in the prototypical perovskite ferroelectrics[5], and the $d^0$ criterion holding for the oxide perovskites that $3d$ transition metal ions must have zero $d$ electrons for ferroelectricity might be avoided.[6]

Among various fluorides, a family of iso-structural materials $Ba_2N_3F_{10}$, $N$=Zn, Co, Ni, have received many interests for their particular crystal structure and complicated magnetic configurations.[7,8] The main obstacle for the studies of fluorides is the harsh conditions for preparing pure phase samples. For example, the traditional method for synthesis of $Ba_2Ni_3F_{10}$ is the solid state reaction, by melting $BaF_2$ and $NiF_2$ in an HF atmosphere sealed in noble metal (such as Au, Pt, etc.) tube at high temperatures (generally above 850 ºC).[7-9]

Till now, most works on fluorides are on bulk samples while very rare fluoride nanostructures were studied.[10,11] In fact, nanostructural materials with reduced dimension are unique important due to their special shape, compositions, chemical and physical properties, which may shed lights to novel optoelectric applications.[12-16] In this work, we develop a novel low-temperature synthesis route based on the hydrothermal method, and pure phase $Ba_2Ni_3F_{10}$ nanowires (NWs) have been successfully prepared under the optimized conditions. Then the optical and magnetic properties of $Ba_2Ni_3F_{10}$ NWs are studied, which show novel three-color fluorescence and exchange bias.

## 2. EXPERIMENTAL AND COMPUTATIONAL METHODS

**2.1. Sample Growth.** Pure phase $Ba_2Ni_3F_{10}$ NWs were synthesized by low temperature hydrothermal method, based on the experience of preparation of pure phase $BaAF_4$ (A=Ni, Mn, Fe, etc.).[17-19] The molar ratio of raw materials $BaF_2$ and $NiF_2$ is 1:4. They were dissolved in trifluoroacetic acid solution (2 ml $CF_3COOH$ and 5 ml distilled water), and the diluted solution was formed by magnetic stirring, which was then put into an 23 mL autoclave. The autoclave was gradually heated to 200 °C, held for 12 or 24 hours, then slowly cooled down to room temperature at a speed of 5~20 °C·h$^{-1}$. The upper remaining liquid was discarded, and the precipitates were kept and washed with ethanol and distilled water for several times. The washed products were placed in a vacuum drying oven and dried at 60 °C, resulting in the final light yellow powders.

**2.2. Characterizations, Optical and Magnetic Measurements.** The structure of $Ba_2Ni_3F_{10}$ NWs was studied by XRD (Rigaku Smartlab3) using a Cu Kα radiation and transmission electron microscope (TEM, Tecnai F20). The morphology was studied by a scanning electron microscope (SEM, FEI Inspection F50). The PL spectra were recorded using a LabRAM HR800 Raman system with 325 nm He-Cd laser excitation. Absorption measurements were carried out with a spectrophotometer (UV-3600) and the fluorescence efficiency (characterized as the absolute quantum yield) was recorded on an Edinburgh FLS920P spectrometer with an integrating sphere. All the optical measurements were conducted at room temperature and atmospheric pressure. The DC magnetization was measured by a SQUID-VSM (SQUID, Quantum Design) from 5 K to 300 K.

**2.3. First Principles Calculations.** The first-principles density function theory (DFT) calculations were based on the projected augmented wave pseudo-potentials using the Vienna ab initio simulation package (VASP).[20-22] The GGA+$U$ method is employed using the Dudarev implementation[4] and various values of the effective Hubbard coefficient ($U_{eff} = U$-$J$) on Ni's $3d$ states have been tested from 0 eV to 2 eV. Both the lattice constants and internal atomic positions are fully optimized until the Hellmann-Feynman forces converged to less than 0.01 eV/Å. The experimental antiferromagnetism was adopted.[7,8] More details of DFT calculation are shown in the Supporting Information.

## 3. RESULTS AND DISCUSSION

**3.1. Fabrication and Morphology of $Ba_2Ni_3F_{10}$ NWs.** In this work, the $Ba_2Ni_3F_{10}$ NWs were prepared by the hydrothermal method at 200 ºC. We only modify the holding time at 200 °C and the cooling rate, and all other parameters are kept identical. Figure 1 (a-j) shows the morphology of $Ba_2Ni_3F_{10}$ samples under various synthesis conditions. The information of samples is denoted as S$n$(B:C), where $n$ is the sample number, B is the holding time ($h$) of reactants at temperature of 200 °C and C represent the cooling rate (°C·h$^{-1}$). It can be seen that all samples exhibit broccoli-like shape. A magnified view can recognize the difference between samples prepared under different conditions. For the sample S1 with the longest holding time and slowest cooling rate, well-shaped Ba$_2$Ni$_3$F$_{10}$ NWs (300 ~ 500 nm in width and several tenth of μm in length) make up of the broccoli. With increasing the cooling rate to 10 °C·h$^{-1}$ (S2) and 20 °C·h$^{-1}$ (S3), the length of NWs is kept, but the width becomes larger (700 ~ 900 nm for S2), and finally the shape changes to microsheets (width of 7 ~ 10 μm for S3). With decreasing the holding time to 12 h, though the NWs morphology is kept, they look rather nonuniform and the surfaces become quite rough, indicating the possible incomplete growth of the NWs. With further increasing the cooling rate to 10 °C·h$^{-1}$ (S5), microsheets instead of NWs are formed. Based on the above observations, we can conclude that slower cooling rate and longer holding time may facilitate the growth of NWs.

All aforementioned samples were characterized by X ray diffraction (XRD), which show similar patterns without any impurity phase, as shown in Figure S1 (supporting information). The typical XRD pattern from sample S1 is shown in Figure 2(a), which matches well with the data of standard PDF NO. 26-0172 of Ba$_2$Ni$_3$F$_{10}$.[9] Selected area electron diffraction (SEAD) and high resolution transmission electron microscopy (HRTEM) were measured on individual single nanowire from sample S1, as shown in Figure 1(k and l). The regularly arranged spots array reveals the single crystalline nature of Ba$_2$Ni$_3$F$_{10}$ individual nanowire. Above characterizations clearly demonstrate the easy and successful fabrication of pure-phase $Ba_2Ni_3F_{10}$. The unit cell of $Ba_2Ni_3F_{10}$ is monoclinic with space group *C2/m*, built up from infinite rutile chains connected together by chains of corner-sharing octahedra and bioctahedral units, as shown in Figure 2(b-c). Ni atoms are distributed among three sublattices and form odd cycles.[7,8] Since sample S1(24:5) exhibits the best nanowire morphologies with pure phase crystalline structure, the following optical and magnetic measurements are mainly performed on this sample.

**3.2. Optical Properties and First-Principles Calculations.** As shown in Figure 3(a), PL spectrum from a bundle of $Ba_2Ni_3F_{10}$ NWs was measured in a range from 340 nm to 800 nm under the excitation of 325 nm laser. Three emission peaks are observed and locate respectively at 360 nm, 530 nm and 700 nm, which occupy the ultra-violet, green and red regions. Such fluorescence is a very promising option for applications in various fields of science and technology.[23-25] There is a sudden jump at around 650 nm, which is an artifact due to frequency multiplication effect of the equipment. To avoid the influence of this artifact, the whole absolute quantum yield ($\eta$) under the excitation of 325 nm was measured and divided into two parts, $\eta$ is 48% in the range of 340 nm ~ 640 nm, while in the range from 670 nm to 750 nm, it is as high as 87%. The peak of 700 nm is much broader than the former two, which suggests that its emission mechanism is probably from defects or vacancies.[26,27] Ultraviolet visible (UV-vis) absorption spectrum of $Ba_2Ni_3F_{10}$ NWs is shown in Figure 3(b). Due to the equipment limitation, the lowest wavelength can only be measured down to 190 nm. Two sharp absorption peaks located at 200 nm and 430 nm, and a broad peak above 550 nm are observed. Obvious absorption edges are seen in P$_1$ and P$_2$, which can be fitted according the Equation (1):

$$\alpha(E) = \frac{A}{E}\left(E - E_{g,dir}\right)^{1/2} + \frac{B}{E}\left(E - E_{g,ind} \mp E_{ph}\right)^2 \quad (1)$$

where α(E) is the absorption coefficient, $E_{g,dir}$ is the direct gap energy, $E_{g,ind}$ is the indirect gap enery, $E_{ph}$ is the phonon energy mediating any indirect gap component, E is the photon energy, A and B are constants.[28] This approach is developed for traditional semiconductors with single parabolic bands and has been extended to analyze other compounds.[29-31] Figure 3(c) and (d) show the linear fitting of (αE)$^2$ vs. Energy, indicating the direct band gap in Ba$_2$Ni$_3$F$_{10}$ NWs. Furthermore, there are three linear regions, which are all applied by the linear fitting. The band gaps obtained by the fitting are 5.4 eV (from the edge of P$_1$), 2.3 and 2.4 eV (from the edge of P$_2$).

To figure out the underline mechanism of such novel emission, we further carried out first-principles calculations. By varying the Hubbard coefficient, it was found that U$_{eff}$=2 eV gives the best description for the Ba$_2$Ni$_3$F$_{10}$. The density of state (DOS) and projected density of states (PDOS) show that the electronic bands near the Fermi level are from the Ni 3d orbitals, as shown in Figure 3(e). Due to the distorted corner- and edge-sharing Ni-F$_6$ octahedra, a weak hybridization between Ni's 3d orbitals and F's 2p orbitals is presented. The local magnetic moment within the Wigner-Seitz sphere was 1.73 μ$_B$ per Ni atom, implying the high-spin state for Ni$^{2+}$ (t$_{2g}^6$e$_g^2$), as expected. Namely, the t$_{2g}$ orbitals are fully occupied and e$_g$ orbitals are half-filled.

The band gap of $Ba_2Ni_3F_{10}$ in our DFT calculation was about 2.37 eV (when $U_{eff}$=2 eV), separating the empty upper-Hubbard $e_g$ bands and occupied lower-Hubbard $e_g$ bands. This Mott gap coincides with the emission phone energy. According to Figure 3(d), the emission and the self-absorption can be well mapped to the transitions among these sub-bands. First, the electrons are excited to the unoccupied upper-Hubbard $e_g$ orbitals of Ni ions and unoccupied $d$ orbitals of Ba ions by the 325 nm laser, leading to the absorption spectra in Figure 3(b), then the excited electrons jump from such unoccupied orbitals to the occupied $e_g$ and $t_{2g}$ orbitals of Ni ions leads to the PL emissions at 360 nm and 530 nm (see Figure 3(a)).

It is rare for a fluorescent material to emit three color light simultaneously and efficiently. Our discovery of $Ba_2Ni_3F_{10}$ as a fluorescent material with interesting three color emissions will shed light to potential applications of optoelectrics. The only drawback is that one color is in the ultraviolet range. In future, the peak position of ultraviolet emission may be shifted to the violet or blue region by the proper ion substitution or structure modification, thus can be applied in the phosphor-converted white light-emitting diodes.

**3.3. Exchange Bias Effect and Possible Multiferroic Material.** Apart from the synthesis and optical properties, magnetic characterization of $Ba_2Ni_3F_{10}$ NWs was also performed. The temperature dependent magnetization (*M-T*) curves measured under a magnetic field of 200 Oe after zero field cooling (ZFC) and field cooling (FC) processes with a cooling field ($H_{cool}$) of 200 Oe. As shown in Figure 4, a splitting of ZFC and FC *M-T* curves is observed below 60 K. Maxima can be observed in both curves, below which magnetization decreases with further decreasing temperature, indicating the antiferromagnetic (AFM) nature. We define the Néel temperature ($T_N$) of 55 K from the maximum of FC $M$-$T$ curve, which is close to the reported $T_N$ of 50 K for single crystal sample.[7] The ZFC and FC $M$-$T$ curves were further measured under a magnetic field of 50 kOe with $H_{cool}$ of 50 kOe for FC process. Clear maximum can still be observed at 55 K, which confirms the AFM transition, and excludes the possible spin glass behavior.[32] Interestingly, the ZFC moment is higher than the FC one in the temperature range close to $T_N$. This abnormal phenomenon might be related to the frustrated spin structure. Competition between strong AFM and weak ferromagnetic (FM) interactions within even cycles of corner-sharing and edge-sharing octahedra leads to a frustrated AFM coupling in bioctahedra units while non-frustrated rutile-like chains exhibit a normal FM coupling.[8] Under the high cooling field, the frustrated spin structure is broken, and there might be some metastable states for the alignment of the spins during the cooling process due to the magnetocrystalline anisotropy. During the warming process, the high field tends to align all the spins to the field direction in both FC and ZFC cases by the Zeeman energy, and slight difference can be observed between the FC and ZFC magnetization with slight larger FC magnetization due to the pre-alignment of spins during the FC cooling process. However, in the temperature range close to $T_N$, the thermal activation is comparable to the AFM interaction. Thus, the spins are easier to be aligned by the measuring field in ZFC case. While in the FC case, the spins are harder to be aligned by the measuring field due to the magnetocrystalline anisotropy, leading to the smaller magnetization in FC case compared to ZFC one.

Figure 5(a) shows the field dependent magnetization (*M-H*) loops of $Ba_2Ni_3F_{10}$ NWs measured at 5 K after cooling from 300 K with $H_{cool}$ of 200 Oe. Clear hysteretic *M-H* loop has been observed, revealing the weak FM behavior.[33-35] It has been reported that non-frustrated rutile-like chains exhibit a normal FM coupling, although previous report achieved only linear *M-H* curves of $Ba_2Ni_3F_{10}$ single crystal at any temperature in the range of 4.2 K ~ 300 K.[7,8] The measuring artifact can be excluded, since we also measured the *M-H* curve at 300 K, and nonhysteretic linear *M-H* curve is observed and shown in the left inset of Figure 5(a). We further measured the *M-H* curve at 20 K and 65 K, which is just below and above $T_N$. The clear hysteresis loop at 20 K and linear curve at 65 K, as shown in Figure S2 (supporting information), further confirming the weak ferromagnetism. Recently, P. Borisov *et al*. have prepared $BaCoF_4$ film epitaxially, and weak ferromagnetism has been observed in contrast to AFM nature of bulk sample, which has been ascribed to anisotropic epitaxial strain.[36] In our case, due to the nanowire shape of our sample with width of 300-500 nm, it is reasonable to assume the core-shell structure, which is commonly adopted to explain the abnormal magnetic phenomena in magnetic nanostructures.[37,38] The core of the nanowire has the perfect crystalline structure, which exhibits frustrated antiferromagnetism with linear *M-H* curve. However, at the shell of nanowire, large strain might exist and the balance between the AFM and FM interactions is broken, and weak ferromagnetism might be realized.[39] Such kind of weak ferromagnetism usually occurs on antiferromagnetic nanostructures.[40,41]

By careful investigation of the *M-H* loop at 5 K with $H_{cool}$ =200 Oe, a clear shift toward the negative field direction has been observed, which is a clear indication of exchange bias (EB).[42] To exclude the possible artifact due to the measuring system, we change the $H_{cool}$ to -200 Oe, and now the *M-H* loop at 5 K shifts to the positive field direction. Furthermore, the *M-H* curves at 300 K passes the original point and shows no shift, which further confirms the validity of observed EB.

Figure 5(b) shows the temperature dependence of exchange bias field ($H_E$) and coercive field ($H_C$). $H_E$ and $H_C$ are defined as $H_E=-(H_L+H_R)/2$ and $H_C=(H_R-H_L)/2$, where $H_L$ and $H_R$ are the left and right coercive fields, respectively. As the temperature increased, $H_E$ gradually decreases to zero at 55 K, which is consistent with the AFM transition temperature. The $H_C$ vs. *T* plot presents a peak value at 55 K with $H_C$ as high as 400 Oe, and then $H_C$ decreases drastically with further increasing temperature. This is a typical behavior of the AFM/FM bilayers,[43] confirming the two layer structure of $Ba_2Ni_3F_{10}$ NWs with AFM core and weak FM shell. Due the interfacial exchange coupling, $H_C$ shows maximum and does not reach zero at the blocking temperature of $H_E$ at 55 K but slightly higher temperature.[43] Besides the shift of *M-H* curve to opposite field direction, a perpendicular shift along the magnetization axis can also be observed.[37] We further measured the temperature dependence of the shift of *M-H* loops along the magnetization axis. Here vertical magnetization shift ($M_{shift}$) and remanent magnetization difference ($M_{rem}$) are defined as $M_{shift}=(M_r^++M_r^-)/2$ and $M_{rem}=(M_r^+-M_r^-)/2$, where $M_r^+$ and $M_r^-$ are the positive and negative remanent magnetization, respectively, indicated in the right inset of Figure 5(a). The vertical shift of loop is a quite commonly observed in exchange bias of those nano magnetic system.[37] Physically, this vertical shift might be due to the uncompensated spins at the interface/surface, which contribute to a significant portion of total magnetization in these systems. Then the pinning of these moments leads to vertical shift in the *M-H* loops.[43] As shown in Figure 5(c), the $M_{shift}$ and $M_{rem}$ show similar temperature dependence as $H_E$ and $H_C$, respectively.

The influence of $H_{cool}$ on $H_E$ and $H_C$ was further studied, and the results are shown in Figure 5(d). $H_E$ increases abruptly with increasing $H_{cool}$ to about 3 kOe and then saturates with further increasing $H_{cool}$, while $H_C$ only increases slightly and then saturates with $H_{cool}$ above 3 kOe. This is significantly distinct from those systems with spin glass layer in which the $H_E$ decreases with increasing $H_{cool}$,[44] and consistent with the typical AFM/FM bilayer system in which $H_E$ increases with increasing cooling field and saturates at high cooling field.[45] This further confirms the core-shell structure of AFM core and weak FM shell in $Ba_2Ni_3F_{10}$ NWs. In order to explore the relationship between maximum applied magnetic field ($H_{max}$) and EB, we measured the *M-H* loops with various $H_{max}$ after cooling under field of 200 Oe. As seen in Figure 5(e) and at *T*=5 K, $H_E$ decreases and $H_C$ increases quickly with increasing $H_{max}$ to 30 kOe, and then both gradually become saturated. This can be understood by that the large maximum field may induce the partial spin flop at the interface between the AFM core and weak FM shell due to the frustrated spin structure of $Ba_2Ni_3F_{10}$.[46]

Although, $Ba_2Ni_3F_{10}$ is not a multiferroic material since its crystal structure is centrosymmetric and no spontaneous polarization is allowed, it is a magnetoelectric material with linear magnetoelectric effect.[35] In bulk $Ba_2Ni_3F_{10}$, and the polarization can be induced under the application of a magnetic field, which has been predicted by Nenert.[47] By preparing $Ba_2Ni_3F_{10}$ in NWs, weak ferromagnetism has been realized. Spontaneous polarization might be introduced along the three crystallographic by the internal magnetic field due to the weak ferromagnetism even without external magnetic field through the linear magnetoelectric coupling. Thus, $Ba_2Ni_3F_{10}$ NWs might be a possible multiferroic material with the coexistence of weak ferromagnetism and ferroelectricity, which needs further studies to confirm. Together with the outstanding optical properties, $Ba_2Ni_3F_{10}$ NWs may become a novel multiferroic material with the simultaneous coexistence and mutual manipulation of optical, magnetic, and ferroelectric properties.

As the single phase multiferroic materials are very rare due to the contradictory requirements of the outer shell electronic structure,[48] our results might shed light on the realization of novel multiferroic materials. By selecting appropriate magnetoelectric materials, weak ferromagnetism might be induced by the proper strain control. Thus spontaneous polarization might be induced through the magnetoelectric effect, leading to the coexistence of ferromagnetism and ferroelectricity.

## 4. CONCLUSIONS

Using hydrothermal technique and under the optimized conditions, we have successfully prepared pure phase $Ba_2Ni_3F_{10}$ NWs. The longer holding time at 200 °C helps the growth of NWs and slower cooling rate narrows the NWs. Under the 325 nm excitation, three color emissions are observed with peak locating in the regions of ultraviolet (360 nm), green (530 nm) and red (700 nm). The ultraviolet and green bands are interpreted to be the electron transitions between the $t_{2g}$ and $e_g$ levels, which is verified by absorption spectra and first-principles calculations, while the broad red emission is suggested to be from defects.

In contrast to the linear *M-H* curve in single crystalline sample, clear hysteresis loop has been observed below the $T_N$, which has been attributed to the induced internal strain at the surface of NWs. EB effect is observed at the temperature below 55 K, and has been explained by the exchange coupling between the AFM core and weak FM shell.

## ASSOCIATED CONTENT

**Supporting Information**

Structural characterization for samples and magnetic data for pure $Ba_2Ni_3F_{10}$ NWs are presented in the supporting information. Figure S1 shown the XRD patterns of the aforementioned five samples synthesized by various conditions. Figure S2 sketched the *M-H* curves measured at $T$ = 20 K and 65 K after cooling with $H_{cool}$ of 200 Oe. The computational method is described in detail.

## AUTHOR INFORMATION

**Corresponding Authors**

* E-mails: Q.X.: xuqingyu@seu.edu.cn; J.D.: jdu@nju.edu.cn; S.D.: sdong@seu.edu.cn

**Author Contributions**

Q.X. conceived and designed the research. S.Z. carried out the experiment. Y.W., Z.W. and S.D. developed the theoretical explanation, J.W. and J.D. carried out the magnetic

measurements, Z.N. contributed to the optical measurement, S.Z. Q.X., and S.D. wrote the paper.

**Note**

The authors declare no competing financial interest.

**ACKNOWLEDGEMENT**

This work is supported by National Basic Research Program of China (No.2014CB921101), the National Natural Science Foundation of China (51172044, 51471085, and 51331004), the Natural Science Foundation of Jiangsu Province of China (BK20151400), and the open research fund of Key Laboratory of MEMS of Ministry of Education, Southeast University.

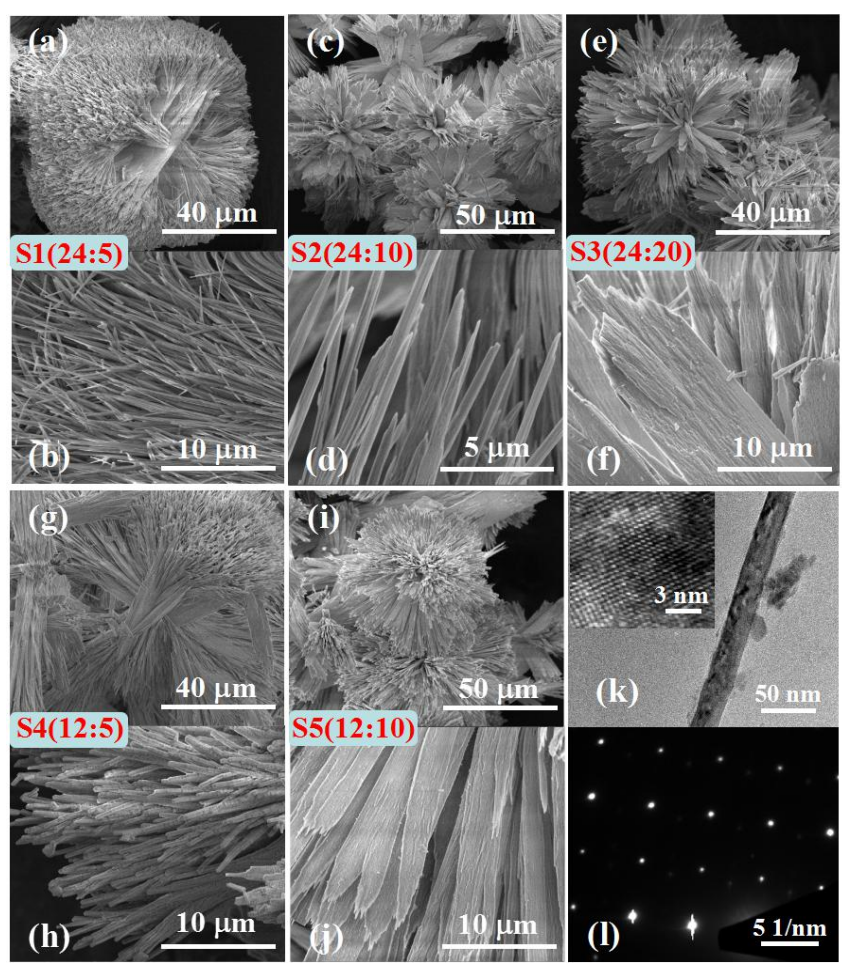

**Figure 1.** (a-j) SEM images of $Ba_2Ni_3F_{10}$ synthesized under different conditions shown in brackets (the left number denotes the holding time at 200 °C in h, and the right number denotes the cooling rate in °C·h$^{-1}$). (k, l) TEM analysis of a single $Ba_2Ni_3F_{10}$ NW. (k) Morphology at a low magnification with the inset showing the corresponding HRTEM image. (l) The SEAD pattern of a single NW.

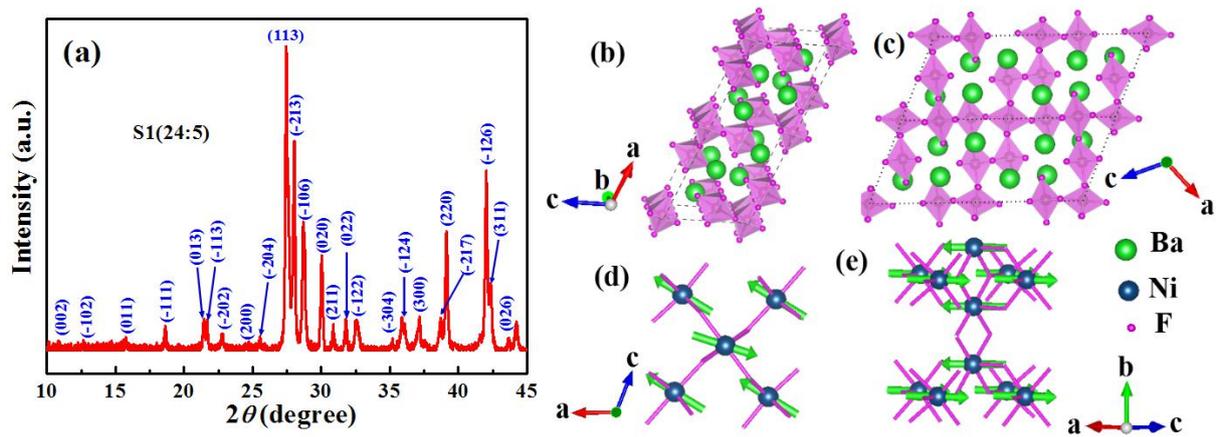

**Figure 2.** (a) XRD pattern of $Ba_2Ni_3F_{10}$ NWs of S1 synthesized using the hydrothermal method. XRD for other samples can be found in the Supporting Information. (b-c) The schematic crystal structure of $Ba_2Ni_3F_{10}$ unit cell. Ba and F atoms are large green and small red spheres, respectively. $NiF_6$ octahedra are plot in light pink. (d-e) The magnetic structure of $Ba_2Ni_3F_{10}$ shown in the ac plane and along the b direction, respectively.

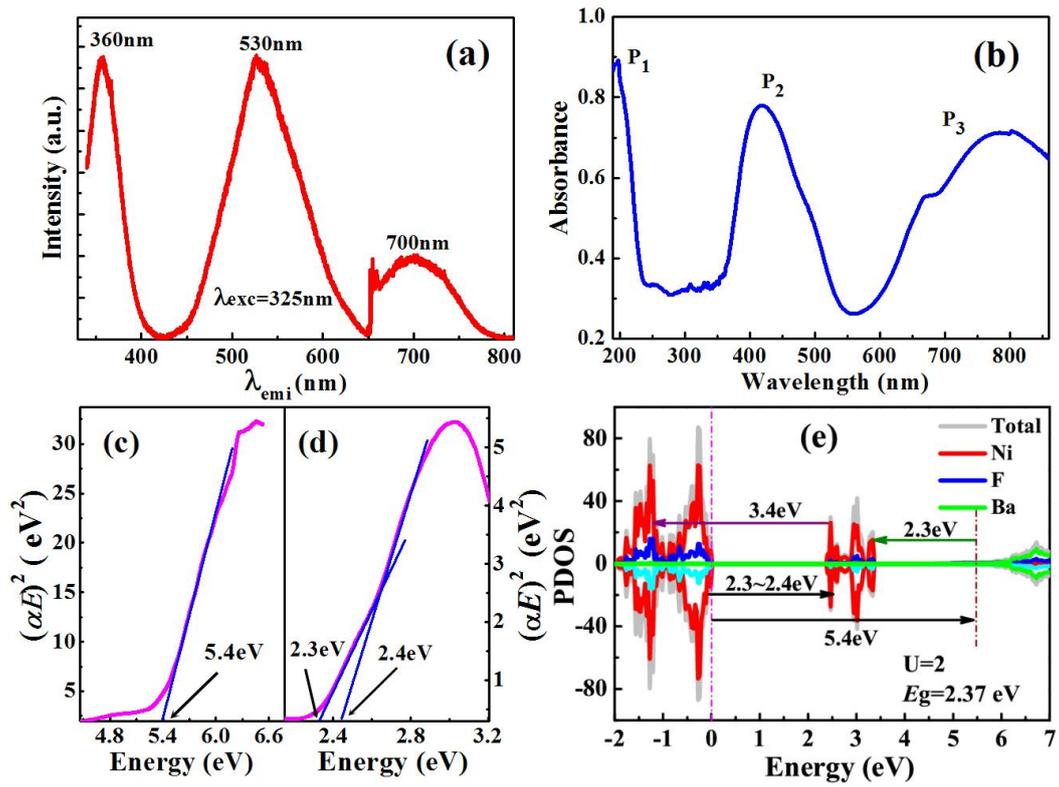

**Figure 3.** (a) PL spectrum of $Ba_2Ni_3F_{10}$ NWs excited by 325nm laser. (b) Absorption spectrum. (c-d) The analysis of band gap from the absorption spectrum. (e) PDOS of $Ba_2Ni_3F_{10}$.

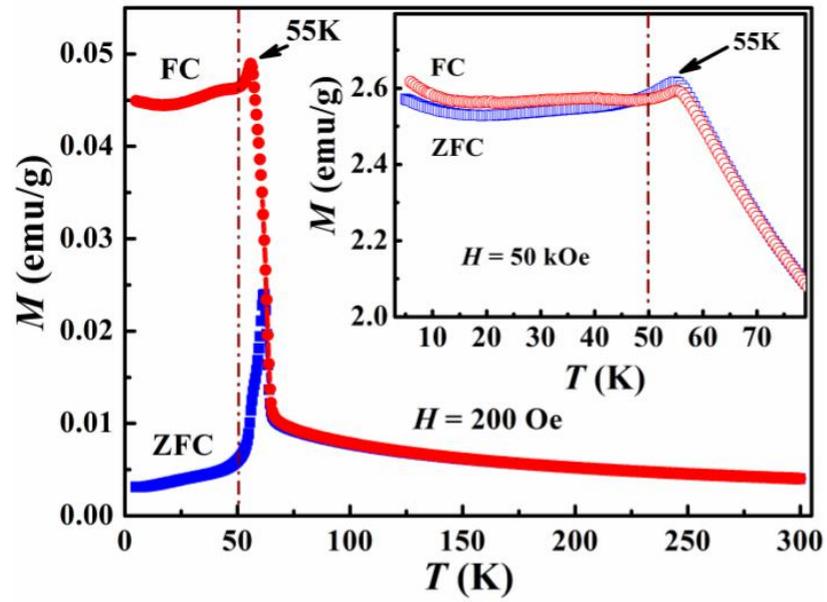

**Figure 4.** ZFC and FC *M-T* curves of Ba$_2$Ni$_3$F$_{10}$ NWs measured under *H* = 200 Oe. ZFC and FC *M-T* curves measured under *H* = 50 kOe (inset).

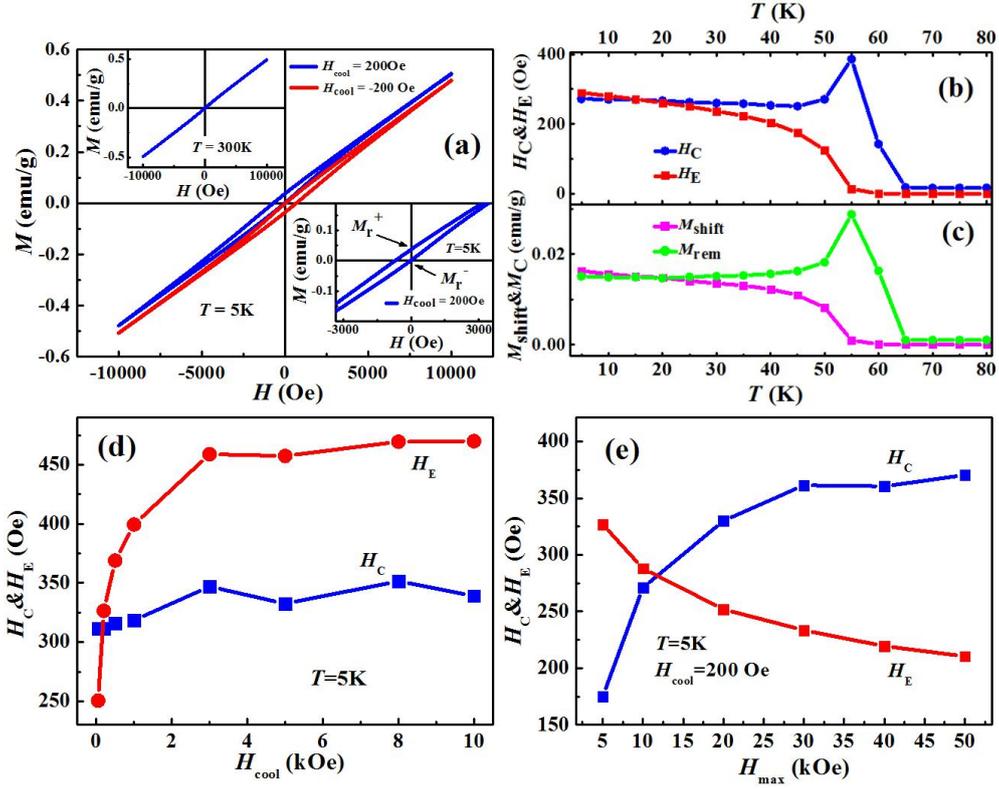

**Figure 5.** (a) *M-H* loops at *T* = 5 K after cooling with $H_{cool}$ of 200 Oe and -200 Oe. Left inset: *M-H* loop measured at *T* = 300 K. Right inset: enlarged view of *M-H* loop of the aforementioned $H_{cool}$ of 200 Oe. (b) *T*-dependence of $H_E$ and $H_C$. (c) *T*-dependence of $M_{shift}$ and $M_{rem}$. (d) Dependence of $H_E$ and $H_C$ on $H_{cool}$ measured at *T* = 5 K with maximum field of 10 kOe. (e) Dependence of $H_E$ and $H_C$ on $H_{max}$ measured at *T*=5 K with $H_{cool}$=200 Oe.

# Table of Contents Graphic

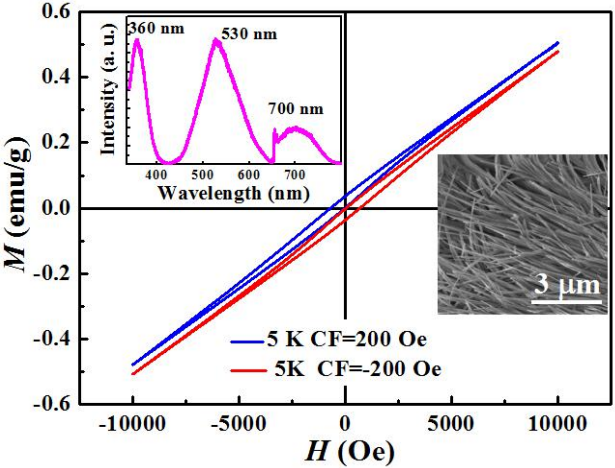

# Supporting Information for

# Synthesis, Optical, and Magnetic Properties of $Ba_2Ni_3F_{10}$ Nanowires

Shuang Zhou[†], Ji Wang[‡,§], Yakui Weng[†], Zhangting Wu[†], Zhenhua Ni[†], Qingyu Xu[*,†,‡], Jun Du[*,‡,§] and Shuai Dong[*†]

[†] Department of Physics, Southeast University, Nanjing 211189, China

[‡] National Laboratory of Solid State Microstructures, Nanjing University, Nanjing 210093, China

[§] Department of Physics and Collaborative Innovation Center of Advanced Microstructures, Nanjing University, Nanjing, 210093, China

Corresponding authors:

* E-mails: Q.X.: xuqingyu@seu.edu.cn; J.D.: jdu@nju.edu.cn; S.D.: sdong@seu.edu.cn

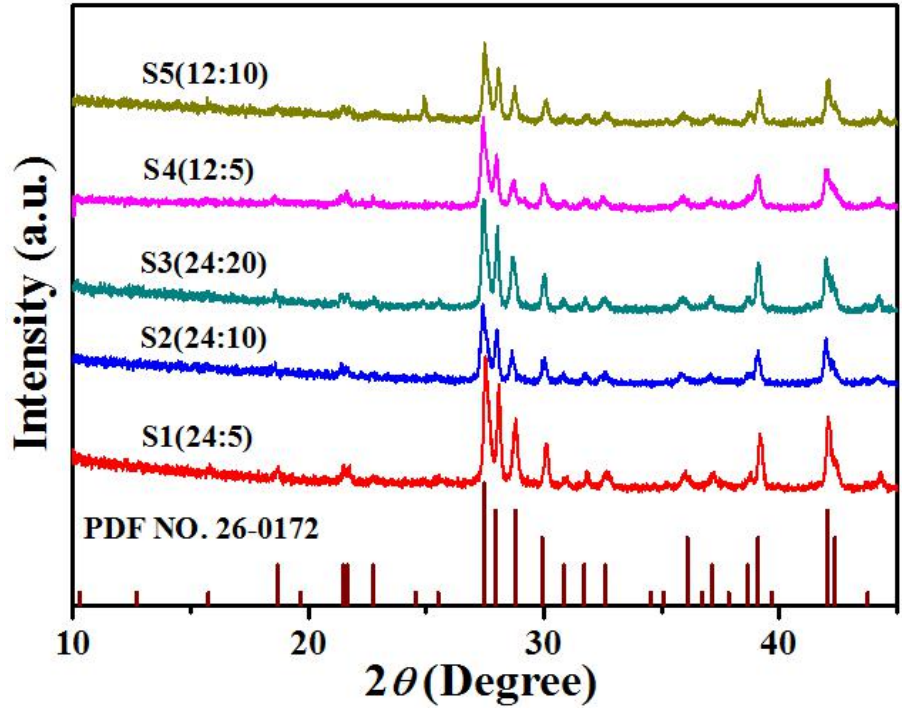

**Figure S1:** XRD patterns of $Ba_2Ni_3F_{10}$ samples with various fabrication conditions.

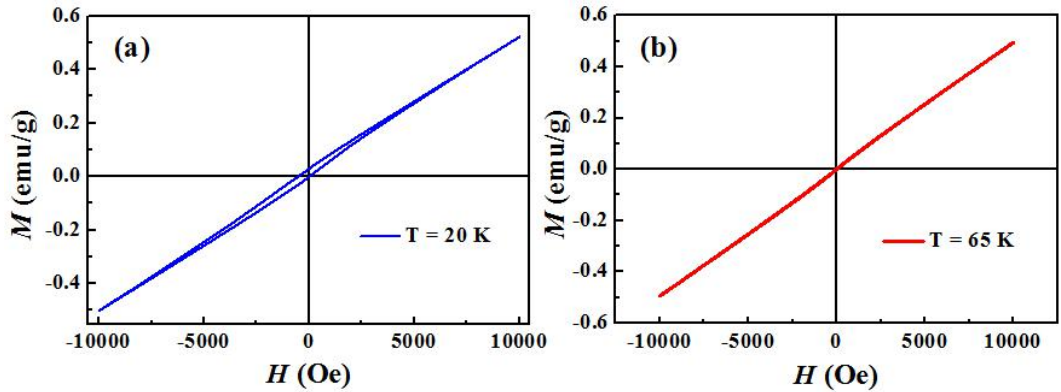

**Figure S2:** (a) *M-H* loops measured at $T = 20$ K after cooling with $H_{cool}$ of 200 Oe, with the $H_C = 207.5$ Oe, and $H_E = 280.5$ Oe. (b) *M-H* loops measured at $T = 65$ K after cooling with $H_{cool}$ of 200 O, and both of the $H_C$ and $H_E$ ~0 Oe.

**First-Principles Calculations.**

The first-principles density function theory (DFT) calculations were based on the projected augmented wave pseudo-potentials using the Vienna *ab initio* simulation package (VASP)[1,2]. The electron interactions are described using PBEsol (Perdew-Burke-Ernzerhof-revised)[3] parameterization of the generalized gradient approximation (GGA). The choice of PBEsol instead of the traditional PBE can give better description of crystalline structure. The cutoff energy of the plane-wave is 500 eV and the Monkhorst-Pack *k*-point mesh is 2x6x5.

It is well known that the DFT calculation underestimates the band gaps, especially for the corrrelated electronic systems with transitional metal ions. To overcome this drawback, the GGA+*U* method is employed using the Dudarev implementation[4] and various values of the effective Hubbard coefficient ($U_{eff} = U-J$) on Ni's *3d* states have been tested from 0 eV to 2 eV. In fact, such a weak Hubbard correction repulsion is enough to give a good description of Ni's *3d* orbitals.

Starting from the experimental values, both the lattice constants and internal atomic positions are fully optimized until the Hellmann-Feynman forces converged to less than 0.01 eV/Å. The experimental antiferromagnetism was adopted,[5,6] i.e. the antiparallel ordering in the *ac* plane and parallel ordering along the *b* direction.